\documentclass[a4paper,twoside]{article}
\usepackage{subfigure}
\usepackage{calc}
\usepackage{amssymb}
\usepackage{amstext}
\usepackage{amsmath}
\usepackage{amscd}
\usepackage{multicol}
\usepackage{pslatex}
\usepackage{apalike}
\usepackage{epsfig}
\usepackage{stmaryrd}
\usepackage[standard]{ntheorem}

\usepackage[table,xcdraw]{xcolor}
\usepackage{cite}
\usepackage[ruled,norelsize]{algorithm2e}
\makeatletter
\newcommand{\removelatexerror}{\let\@latex@error\@gobble}
\makeatother
\usepackage{array}
\usepackage{algpseudocode}
\usepackage{pifont}
\usepackage{stmaryrd}
\usepackage{mwe}
\usepackage{mdwmath}
\usepackage{mdwtab}
\usepackage{eqparbox}
\usepackage{url}
\usepackage{enumitem}
\usepackage{amsfonts}
\usepackage{makeidx}
\usepackage{epstopdf}
\usepackage{multirow}
\usepackage{alltt}
\usepackage{dsfont}

\usepackage{adjustbox}

\usepackage{amssymb}

\newcolumntype{L}[1]{>{\raggedright\let\newline\\\arraybackslash\hspace{0pt}}m{#1}}
\newcolumntype{C}[1]{>{\centering\let\newline\\\arraybackslash\hspace{0pt}}m{#1}}
\newcolumntype{R}[1]{>{\raggedleft\let\newline\\\arraybackslash\hspace{0pt}}m{#1}}

\begin{document}

\title{One random jump and one permutation: sufficient conditions to 
chaotic, statistically faultless, and large throughput PRNG for FPGA}

\author{Mohammed Bakiri, Jean-Fran\c cois Couchot, and Christophe Guyeux}


\maketitle
\abstract{Sub-categories of mathematical topology, like the mathematical theory of chaos, offer interesting applications devoted to information security. In this research work, we have introduced a new chaos-based pseudorandom number generator implemented in FPGA, which is mainly based on the deletion of a Hamilton cycle within the $n$-cube (or on the vectorial negation), plus one single permutation. By doing so, we produce a kind of post-treatment on hardware pseudorandom generators, but the obtained generator has usually a better statistical profile than its input, while running at a similar speed. We tested 6 combinations of Boolean functions and strategies that all achieve to pass the most stringent TestU01 battery of tests. This generation can reach a throughput/latency ratio equal to 6.7 Gbps, being thus the second fastest FPGA generator that can pass TestU01. 
}

\section{\uppercase{Introduction}}
\label{sec:introduction}
The theory of chaos, which refers here to a sub-category of the mathematical topology discipline, proposes attractive applications in computer science, as in the information security field. For instance, various authors propose to use chaos for pseudorandom numbers generation (PRNGs, which are not true random number TRNGs, the former being deterministic algorithms based on iterative processes like recurrent sequences, while the latter use a physical or mechanical phenomenon as sources of randomness), leading to the notion of chaotic pseudorandom number generators (CPRNGs). Rigorously speaking, CPRNGs are non-linear algorithms of the form $x_0 \in \mathbb{R}$: $x^{t+1} = f(x^t)$, where $f$ is a chaotic continuous map on a given topological space, and $x^t$ is the $t$-th term of a sequence $x$. In practice, and unfortunately, this acronym often improperly refers to any attempt to use an element of chaos (logistic map, etc.) within an algorithm, and for pseudorandom generation purpose. 

Reasons explaining the use of chaos for randomness generation encompass their sensitivity to initial conditions, their unpredictability, and their ability of reciprocal synchronization~\cite{pecora1990synchronization}. However, due to finite precision and quantization of floating point numbers, a CPRNG may exhibit both deflated periods and non uniformly distributed outputs when designing it on finite state machines. Additionally, from a security point of view, these chaotic PRNGs have most of the times major drawbacks that are frequently reported~\cite{wiggins2003introduction}.
To face these drawbacks, an original work was firstly introduced in~\cite{bgw09:ip, GuyeuxThese10}, in which the authors present a new way to generate pseudorandom numbers that rigorously satisfies the mathematical properties of chaos, as defined by Devaney~\cite{Devaney},  Li-Yorke~\cite{Li75}, and so on. The main idea is to only manipulate bounded integers, while iterating on an infinite countable set. By doing so, what is designed on computers is exactly what is studied theoretically, the whole algorithm is mathematically proven as chaotic, and they do not deal with floating point numbers.


Following this approach, this article investigates the hardware point of view, and targets to reach speed generation with good statistical profile, while being chaotic. Contributions can be summarized as follows. A new pseudorandom number generator, specifically designed for FPGA, is proposed. At each iteration, it receives a new input from another given generator, called the strategy. Thanks to an embedded Boolean function and a permutation, our generator, which can be considered as a post-treatment on the inputted one, has usually a better statistical profile than its input, while running at a similar speed. With more details, we tested 6 combinations of functions and strategies that all achieve to pass the whole TestU01, which is the most stringent battery of tests currently available. This generation can be achieved with a throughput/latency ratio equal to 6.7 Gbps, which is close to the fastest existing FPGA generator (that can pass TestU01 as XOR-CIPRNG). 
The proposal has been fully deployed on a FPGA, it runs completely in parallel while consuming as low resources as possible.

The remainder of this article is organized as follows. The next section recalls various proposals in the use of chaos for hardware pseudorandom number generation. Their FPGA implementation is presented, explaining how to compare them in terms of hardware resources and statistical behavior. Section~\ref{sec:proposal} describes our proposed design for a new chaotic PRNG, targeting a FPGA implementation.
Then, in Section~\ref{sec:CIG-FPGA}, the hardware platform used to evaluate all evoked chaotic PRNGs is presented. Statistical comparisons are provided in the same section, using the well-known TestU01 battery of tests. Finally, this article ends by a conclusion section, in which the contribution is recalled and intended future work is outlined.

\section{\uppercase{Chaotic Random Number Generators}}
\label{sec:prng}

This section first presents an exhaustive list of PRNGs that are linked to a chaotic behavior in one way or another. 
It next presents their FPGA implementation to compare them in terms of
hardware resources and statistical behavior. 

\subsection{Chaotic PRNGs}
\label{relwork}


\textit{Chaotic Mapping PRNG.} 
Most of these generators are based on the \emph{Logistic Chaotic Map}, also called the ``LCG'' map~\cite{may1976simple}, defined as follows: $x^{t+1} = r \times x^t(1-x^t)$, where $0<x^{t}<1$ and $r$ is the \emph{biotic potential} $(3.57<r<4.0)$. 
The logistic map mainly depends on the parameter $r$: its chaotic behavior is lost when $r$ is out of the range provided above. 
The second most frequently used function is the \emph{H\'enon chaotic map}~\cite{henon1976two}, which takes a point ($x^t$,$y^t$) within the plan square unit and maps it into a new point ($x^{t+1}$, $y^{t+1}$). This map is defined by these equations: $x^{t+1} = (1-a (x^t)^2)+y^t$ and $y^{t+1}= bx^{t+1}$, where $a$ and $b$ are called \emph{canonical parameters}.

In~\cite{dabal2011chaos}, the authors have used fixed point representation~\cite{padgett2009fixed} to implement the logistic map using Matlab DSP System Toolbox software. 
They generate many designs with different lengths from $16$ to $64$ bits, where the resources are depending on the precision ($24$ to $53$ bits). 
Authors of~\cite{dabal2012fpga} compare this implementation with another chaotic PRNG based on the H\'enon map. 
Unlike the logistic map, the $64$ bits multiplication in H\'enon map cannot be implemented with a left shift operation, which leads to the use of DSPs blocks of the FPGA for all multiplications needed to implement $a(x^t)^2$. 
Two optimized versions of PRNGs based on chaotic logistic map are proposed in~\cite{6868789}, which aim to reduce resources and increase frequency.
The objective of these two PRNGs is to pipeline the multiplication operations and synchronize them while adding some delays into each stage, in order to ensure a parallel execution of sequences. 

In~\cite{pande2010design}, the authors vary the biotic potential $r$ and observe the divergence of random for almost all initial values. 
Accordingly, they propose a range of the form $[\alpha, 1-\alpha]$, where $\alpha < 0.5$. 
Another way to select the parameter $r$ is presented in~\cite{liu2008improved}. They propose a couple of two logistic map PRNGs, each having different seed and parameter ($x_0$, $r_1$ and $y_0$, $r_2$ respectively), where both generates pseudorandom numbers synchronously. The main idea is to recycle the pseudorandom number generated by the first chaotic map, namely $x^{t+1}$, as the biotic potential $r_2$ for the second one ($y^{t+1}$) when either $3.57<x^{t+1}<4$ is satisfied or the sequence output is divergent.

Finally, in~\cite{giard2012fpga} four different chaotic maps are implemented in FPGA, namely, the so-called \emph{Bernoulli}, \emph{Chebychev}, \emph{Tent}, and \emph{Cubic} chaotic maps. The implementation is done with and without FPGA's DSP blocks for the multiplication operations. The results show that the Bernoulli chaotic map gives a higher ratio of area/power compared to the other chaotic proposed generators.

\textit{Chaotic based Timing Reseeding (CTR).} Authors of~\cite{vcernak1996digital} address the short period problem due to the quantization error from a nonlinear chaotic map PRNG. Instead of initializing the chaotic PRNG with a new seed, the seed can be selected by masking the current state $x^{t+1}$ at a specific time. 
More precisely, the reseeding unit compares the two register states to check whether a fixed point has been reached. 
This main concept of CTR was first implemented in FPGA~\cite{li2006chaos}, in which the \emph{Carry Lookahead Adder} has been used to optimise the critical path of the partial products of the multiplication operation. Authors of~\cite{li2012period} present more hardware details for reducing multiplication operation resources. They also mix the output from the PRNG with an auxiliary generator $y^{t+1}$ to improve statistical tests. 
Finally, they suggest to choose a reseeding period that must be not only prime, but also not a multiple of the nonlinear chaotic map PRNG. 

\textit{Differential Chaotic PRNG.} This is a digitized implementation of a nonlinear chaotic oscillator system in R\"ossler format~\cite{Rössler1976397}.
It uses an approximated numerical solution to solve a generalization of the L\"orenz hyperchaos equation.
The resolution 
was the main study done in~\cite{zidan2011effect} with other differential systems as the Chen and Elwakil ones.
The authors design 
various numerical methods for each system. 
They 
show that obtained results with the Euler numerical approach are the best regarding area and throughput perspectives.

More details regarding implementation and optimization of this L\"orenz Equation
are given in~\cite{zidan2011random}. In this article, authors proposed to use again an Euler approximation  with less area but same range of throughput.
Authors of~\cite{dabal2012fpga}, for their part, have implemented the so-called \emph{Oscillator Frequency Dependent Negative Resistors} (OFDNR)~\cite{elwakil2000chaotic},
and use the same Euler approximation.

\textit{Chaotic Iteration based PRNG (CIPRNG).} 
Formally speaking, this is a random walk in the graph of iterations of a specific binary function.
The direction to take and the path length are defined by the embedded generator(s). Practically, it can be seen as post processing treatment which adds chaos (as defined by Devaney) to the embedded PRNG~\cite{bg10:ip}. 
A first application of such an approach was presented in the PRNG framework, leading to the so-called chaotic iterations based pseudorandom number generators (CIPRNG,~\cite{fang2014fpga,bahi2013fpga}). Since then, various improved versions have been proposed, one of them being designed specifically for FPGAs. This latter has been recently updated, in which two CIPRNG variants for FPGA have been designed, namely the XOR-CIPRNG and the CIPRNG-MultiCycle, see~\cite{secrypt16}. 

Meanwhile, 
in~\cite{couchot2014pseudorandom}, the authors have proposed to remove a \emph{Hamilton Cycle}, satisfying some balance properties, from the Markov chain on the $n$-cube, while  
in~\cite{contassot2017random}, authors 
proposed new functions without a Hamilton Cycle, and studied the length of the walk in their cube, until having an associated \emph{Markov} graph close enough to the uniform distribution. 
In these first studies, the minimum  length of the chain between two uniform outputs is larger than $109$, which id prohibitive.

\subsection{FPGA Performance Analysis}
In this section, Table~\ref{Tab:fpgaimp1} and Table~\ref{Tab:fpgaimp2} 
resume the implementation results of different chaotic PRNG presented in previous Section~\ref{relwork} 
and recall results of linear PRNGs (already presented in~\cite{secrypt16} which pass TestU01). 

In order to compare the considered generators, we thus have computed throughput (rate), latency, and the ratio throughput/latency of both. Obtained results on FPGAs are provided in Tables~\ref{Tab:fpgaimp1} and~\ref{Tab:fpgaimp2}. We can conclude that the three best chaotic generators for FPGA are, namely, the one from~\cite{6868789} that uses the logistic map with Matlab simulink macros, the chaotic iterations based PRNG~\cite{secrypt16}, and the one based on the chaotic \emph{Bernoulli} map~\cite{giard2012fpga}. If we consider the linear PRNGs who pass TestU01 (see section below), they have the worst throughput due to their use of multiplications and their various dependencies.
However, to have a large throughput does not mean to produce an uniform distribution of numbers, which leads to the investigation of statistical results.

\subsection{Statistical Tests}
Statistical tests are fast methods to study in practice the randomness of generated numbers, by the mean of software batteries.
They are based on various mathematical and physical approaches, and are thus used as generator benchmarks. 
To perform comparisons, in this study, we considered the reputed NIST SP$800-22$~\cite{NIST:2010:Online} and TestU01~\cite{l2007testu01} batteries of tests. On the one hand, the NIST battery, 
considers 15 tests to evaluate the randomness of a sequence of fixed length $10^6$, where each test is passed if its $p$-value is larger than $0.0001$. 
On the other hand, TestU01~\cite{l2007testu01} is the most complete and stringent battery to pass, which embeds 7 big batteries that test more than $10^{32}$ pseudorandom values for each inputted generator.
TestU01 consider passing tests if the $p$-value is $ \in[0.001,0.999]$.

From our first experiments, it can already be noticed that almost all chaotic PRNGs can pass the NIST batteries, but they fail on TestU01, with the exception of XOR-CIPRNG. Conversely, some linear PRNGs like PCG32, xorshift64*, or MRG32, chosen for comparison in this study, can also pass the TestU01.
This is why the work in~\cite{6868789} and in~\cite{giard2012fpga}, based on the logistic map and the Bernoulli one, will be used for throughput comparison, while linear PRNGs will be considered for statistical tests.
We can however already conclude that only XOR-CIPRNG satisfies both low hardware resources and a success against the TestU01 battery, which has already been stated in~\cite{secrypt16}.

\begin{table*}[ht]
\centering
\caption{FPGA implementations of chaotic PRNGs}
\label{Tab:fpgaimp1}
\begin{adjustbox}{max width=\textwidth}
\def\arraystretch{1.3}
\begin{tabular}{|l|c|c|c|c|c|c|}
\hline
\text{PRNG} & \multicolumn{6}{c|}{\text{32 Bit Chaotic}} \\ \hline
\text{} & \multicolumn{1}{l|}{\text{\cite{secrypt16}}} & \text{\cite{dabal2011chaos}} & \text{\cite{dabal2012fpga}} & \text{\cite{6868789}} &  \text{\cite{li2006chaos}} & \text{\cite{li2012period}} \\ \hline
\text{Function} & XOR-CIPRNG {[}A,B,2{]} & LCGM & \begin{tabular}[c]{@{}c@{}}LCG-\ H$\acute{e}$non-\ FNDR\end{tabular} & LCG  &\begin{tabular}[c]{@{}c@{}} Timing\ Reseeding\end{tabular} & \begin{tabular}[c]{@{}c@{}}Timing\ Reseeding\end{tabular}  \\ \hline
\text{Frequency (Mhz)} & 258 & 76.1 & 151.1\ -\ 58.2\ -\ 183 & 233  & 200 & 200 \\ \hline
\text{DSP} & 0 & 4 & 4-4-0 & 16 & 0 & 0 \\ \hline
\text{Area} &  7568  & 784 & 640\ -\ 4568\ -\ 4568 & 9240 & *** & 11903 \\ \hline
\text{Design Latency} & 2 & *** & *** & 8\ to\ 16  & *** & *** \\ \hline
\text{Output Latency} & 1 & 1 & 1 & 1 & 1 & 1 \\ \hline
\text{Throughput/Latency (Gbps)} & 8.30 & 2.435 & 4.835\ -\ 1.862\ -\ 5.856 & 7.5 & 6.4 & 6.4 \\ \hline
\text{NIST Test} & PASS & PASS & PASS & PASS  & PASS & PASS \\ \hline
\text{TestU01 (BigCrush)} & PASS & NO & NO & NO  & NO & NO \\ \hline
\end{tabular}
\end{adjustbox}
\end{table*}

\begin{table*}[ht]
\centering
\caption{FPGA implementation of chaotic (continuation of Table~\ref{Tab:fpgaimp1}) and linear PRNGs}
\label{Tab:fpgaimp2}
\begin{adjustbox}{max width=\textwidth}
\def\arraystretch{1.1}
\begin{tabular}{|l|c|c|c|c|c|c|c|}
\hline
\text{PRNG} & \multicolumn{3}{c|}{\text{32 Bit Chaotic}} & \multicolumn{4}{c|}{\text{Linear}} \\ \hline
\text{} & \multicolumn{1}{l|}{\text{\cite{zidan2011effect}}} & \multicolumn{1}{l|}{\text{\cite{fang2014fpga}}} &  \multicolumn{1}{l|}{\text{\cite{giard2012fpga}}} & \multicolumn{1}{l|}{\text{PCG32}} & \multicolumn{1}{l|}{\text{xorshift64*}} & \multicolumn{1}{l|}{\text{KISS32}} & \multicolumn{1}{l|}{\text{MRG32}} \\ \hline
\text{Function} & \begin{tabular}[c]{@{}c@{}}LRZ\ -\ Chen\ -\ ELW\end{tabular} & ICPRNG & B\ -\ CH\ -\ T & LCG & xorshift & Combine & Combine \\ \hline
\text{Frequency (Mhz)} & 53.53\ -\ 122\ -\ 126.7 & 200 & 265.9\ -\ 118.7\ -\ 111.8 & 112 & 113 & 100 & 106 \\ \hline
\text{DSP} & 8\ -\ 0-\ 0 & *** & 0 & 0 & 0 & 0 & 0 \\ \hline
\text{Area} & 3064\ -\ 13968 & 3652 & *** & 27632 & 27968 & 26520 & 33456 \\ \hline
\text{Design Latency} & *** & *** & *** & 17 & 21 & 9 & 14 \\ \hline
\text{Output Latency} & 1 & 1 & 1 & 17 & 21 & 9 & 14 \\ \hline
\text{Throughput/Latency (Gbps)} & 1.71\ -\ 3.9\ -\ 4.06 & 6.4 & 8.5\ -\ 3.798\ -\ 3.577 & 0.189 & 0.34 & 0.8 & 0.24 \\ \hline
\text{NIST Test} & NO & PASS & PASS & PASS & PASS & PASS & PASS \\ \hline
\text{TestU01 (BigCrush)} & NO & PASS & NO & PASS & PASS & NO & PASS \\ \hline
\end{tabular}
\end{adjustbox}
\end{table*}

\section{\uppercase{The proposal}}
\label{sec:proposal}
The previous section ends with the idea that it is hard to have together the three properties of: chaos, hardware efficiency, and a random-like statistical profile.

\subsection{General idea}
\begin{figure}[!h]
    \centering
    \includegraphics[width=\columnwidth]{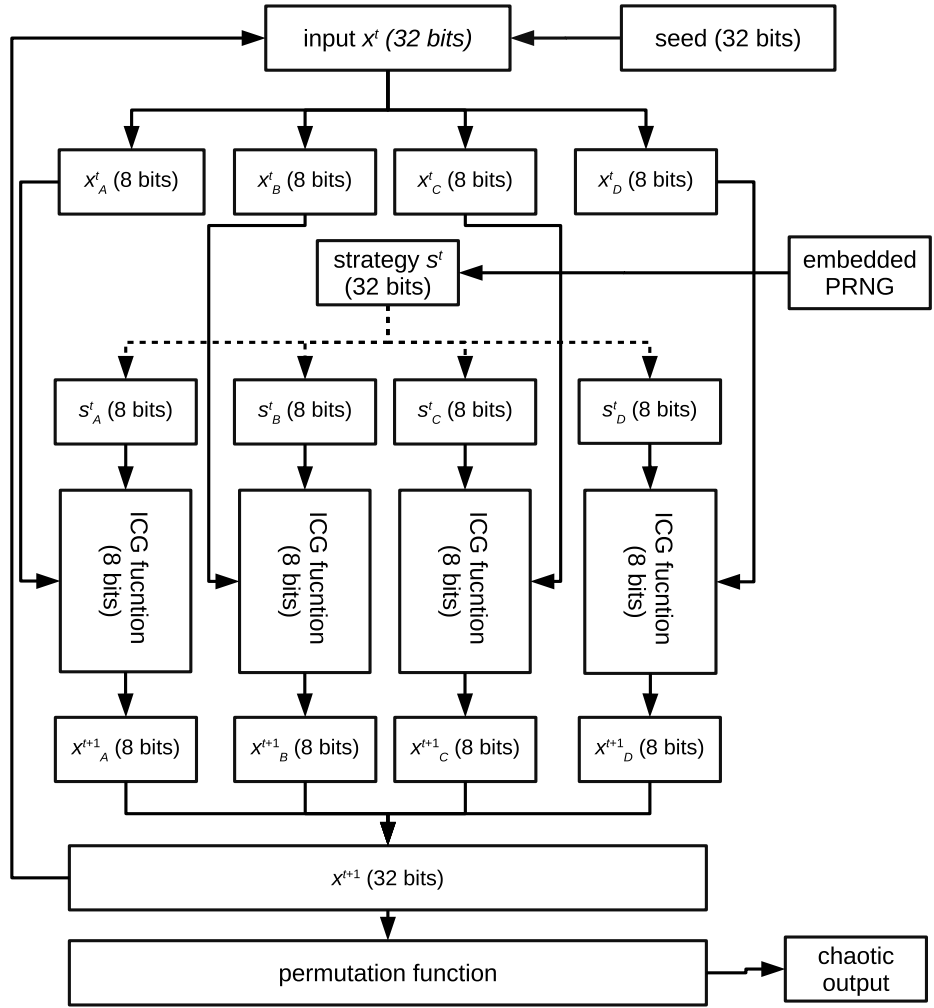}
    \caption{The proposal}
    \label{fig:proposal}
\end{figure}

Let us first discuss on how we tackle this problem.
The first key idea is to have a short internal state, possibly split into parallel blocs. 
This divide and conquer approach aims at ensuring hardware efficiency but is in conflict with statistical quality.  
Chaotic iterations~\cite{bahi2013fpga,fang2014fpga} can be used to achieve chaos objectives.
However, 
the general formulation of the chaotic iterations~\cite{bcgh15:ij} should be preferred than the original one when efficiency is needed. 
Finally, permutation techniques~\cite{citation-0} have presented a convenient way to ensure statistical faultless, in a fast manner.
Our proposal is based on these three main ideas and is summarized in Figure~\ref{fig:proposal}. 

At first, it can be seen that the seed $x^0$, the internal state $x^t$,  and the output $x^{t+1}$ are all expressed with the same number $\mathsf{N}$ of bits. Without loss of generality, we consider hereafter that $\mathsf{N}=32$.
Let us show how to produce a new output $x^{t+1}$ for a given input $x^t$.
This one is first split into $n$ blocs of equal length. 
We consider here  that $n=4$ and we thus have $x^t= (x^t_A, x^t_B, x^t_C, x^t_D)$ where 
$x_l^t$ is of size 8 for $l \in\{A,B,C,D\}$.
The next step consists in obtaining a $\mathsf{N}$-bits number $s^t$ from another embedded PRNG, which is called \textit{the strategy}.
Similarly to $x^t$, the vector $s^t$ is  split into $n$ blocs. Here we thus obtain $s^t= (s^t_A, s^t_B, s^t_C, s^t_D)$. Each $s_l$, $l \in\{A,B,C,D\}$, can be interpreted as a set of elements in $\{1,2,\dots, 8\}$.
Each bloc $x^t_l$ is modified separately as the result of the general formulation of the Chaotic Iterations~\cite{bcgh15:ij} applied on $x^t_l$, $s^t_l$ and a specific function $f: \mathbb{B}^8 \to \mathbb{B}^8$, as described hereafter. The $i$-th component of $x^{t+1}_l$ is the $i$-th one of $f(x^t_l)$ if $i$ is within the set $s^t_l$, else this component is the $i$-th one of $x^t_l$ (\textit{i.e.}, only the components indicated by the set $s^t_l$ are updated).
This results $x^{t+1}_l$. 
All the $x^{t+1}_l$ are concatenated hereafter, producing the new internal state  $x^{t+1}$.
Finally, a permutation over the $\mathsf{N}$ bits is applied on $x^{t+1}$ to produce the new output.

The choice of the function $f$ executed  inside the ICG iteration, of the embedded PRNG, and of the chosen final permutation function has a great influence on the quality of the generator. It is discussed in the next sections.

\subsection{Iterated Function}

Let 
$s \in  (\mathbb{B}^8)^{\mathbb{N}}$ be a sequence of subests of $ \{ 1, \dots \mathsf{N}\}$,  
$x^0$ be a vector in $\mathbb{B}^8$, and 
$f$ be a function from $ \mathbb{B}^8$ to $\mathbb{B}^8$. 
The sequence $(x^t)^{t \in \mathbb{N}}$ of vectors in $\mathbb{B}^8$ defined according 
the general formulation of the Chaotic Iterations~\cite{bcgh15:ij} is
$$
x^{t+1} = (x^{t+1}_1, \dots, x^{t+1}_{\mathsf{N}}) \textrm{ where } 
x^{t+1}_i = \left\{ 
\begin{array}{l}
f_i(x^t)  \textrm{ if $i \in s^t$,}\\
x_i^t  \textrm{ otherwise.}\
\end{array}
\right.
$$

Two functions from $\mathbb{B}^8$ to $\mathbb{B}^8$ are mainly studied in this article.
The former is the negation function, further denoted as $\text{NEG}$.
In this one, each $f_i$ is defined with $f_i(x) = \overline{x_i}$. 
For instance, the image of 5 = 00000101 is 250 = 11111010.
The latter, denoted as F1,  is a function whose graph of generalized iterations is strongly connected and 
which has been obtained by removing a balanced Hamiltonian cycle in a
$\mathsf{N}$-cube following the method suggested in~\cite{contassot2017random}. 
These two functions are recalled in Table~\ref{tab:boolean}.
The choice of these two functions is motivated by the objective to make
the iterations chaotic.

\begin{table*}[ht]
\centering
\caption{Boolean functions}
\label{tab:boolean}
\begin{tiny}
\begin{tabular}{|l|c|}
\hline
\text{Function} & \text{$f(x)\ for\ x \in  [0, 1, 2, 3, 4, 5 \dots , 2^{n}-1]$} \\ \hline
 \text{NEG} & \begin{tabular}[c]{@{}l@{}}
 [255, 254, 253, 252, 251, 250 \dots, 9, 8, 7, 6, 5, 4, 3, 2, 1, 0] \end{tabular}   \\ \hline
\text{F1} & 
\begin{tabular}[c]{@{}l@{}}
[223, 190, 249, 236, 243, 234, 241, 252,  183, 244, 229, 245, 179, 178, 225, 248,  237, 254, 173, 232, 171, 202, 201, 200,\\  
247, 198, 228, 230, 195, 242, 233, 160,  215, 220, 205, 216, 218, 154, 221, 208,  213, 210, 212, 148, 147, 211, 217, 209,\\  
239, 238, 141, 140, 235, 203, 193, 204,  135, 134, 199, 197, 131, 226, 129, 224,  63, 174, 253, 184, 251, 250, 189, 176, \\ 
191, 246, 180, 182, 51, 50, 185, 240, 47,  46, 175, 188, 139, 42, 161, 172, 231, 164,  181, 165, 227, 130, 33, 32, 31, 222, 153,\\  
158, 219, 26, 25, 156, 159, 214, 151, 149,  146, 18, 144, 152, 207, 206, 157, 136,  138, 170, 169, 8, 133, 6, 5, 196, 3, 194,\\  
137, 192, 255, 110, 109, 120, 107, 126,  125, 112, 103, 114, 116, 118, 123, 98,  121, 96, 79, 78, 111, 124, 75, 122, 97, \\ 
108, 71, 100, 117, 101, 115, 66, 113, 64,  127, 90, 89, 94, 83, 91, 81, 92, 95, 84,  87, 85, 82, 86, 80, 88, 77, 76, 93, 72, 74,\\  
106, 105, 104, 69, 102, 68, 70, 99, 67, 73,  65, 55, 58, 57, 44, 187, 186, 49, 60, 119,  52, 37, 53, 35, 54, 177, 56, 45, 62, 61,\\  
40, 59, 10, 9, 168, 167, 166, 36, 38,  163, 162, 41, 48, 23, 28, 13, 24, 155,  30, 29, 16, 21, 150, 20, 22, 27, 19, 145,\\  
17, 143, 142, 15, 14, 43, 11, 1, 12, 39,  4, 7, 132, 2, 34, 0, 128]
\end{tabular} 
\\ \hline
\end{tabular}
\end{tiny}
\end{table*}

\subsection{Permutation Function}

First of all, our proposal is a parallel execution of 4 blocks, each one producing 8 bits. The internal state $x$ is next produced as the concatenation of the results of the 4 blocks.
This design is guided by the goal of reducing the required resources. However, such an approach
suffers from decreasing the statistical complexity of the PRNG: without any post treatment it would be dramatic, because it is equivalent to deal with 8 bits only. 
A final step which scrambles the internal state is thus necessary to tackle this problem. 

This can be practically implemented with a permutation function 
(which allows to obtain a uniform output) provided  it does not break the chaos property (as proven in the next section).
Among the large choice of permutation functions (such as rotation, dropping, xoring...),
we inspire from one detailed in~\cite{citation-0}. This work indeed proposes a bench of permutation functions allowing to succeed statistical tests. 


%
%
This permutation function is implemented as in Algorithms~\ref{eq:perm}. 
It is not hard to see that it is mainly a composition of three subfunctions.
Let \textit{In32} be the internal state. 
The first function scrambles  between 17 and 28 rightmost bits (\textit{i.e.} middle bits) 
with a xor function. The number of selected elements depends on the value of $\textit{In}32$.
Then, the second function applies a modular multiplication in the cyclic group of elements in $\{1, \dots, 2^{31}-2 \}$. 
The chosen multiplier $b$ 
is a primitive root of the modulus $2^{31}-1$. However, in~\cite{citation-0} they need more than 36 bits of internal state to pass TestU01, which is equivalent of a modulus of $2^{37}-25$. Therefore,
$b$ is set to 277803737, but any primitive root of $2^{37}-25$ is convenient for their work in~\cite{citation-0}.
The latter function is a simple right xorshift on the lowest bits to scramble them.

\begin{figure}[ht]
\begin{small}
\centering
\removelatexerror
  \begin{algorithm}[H]
  \DontPrintSemicolon
\caption{Random Xorshift Permutation Function}
\label{eq:perm}
\KwIn{$In32$  32-bit word)}
    \KwOut{$Out32$ (a 32-bit word)} 
    \ \ \ \ $word1 \gets (In32 \gg ((In32 \gg 28u) + 4u)) \otimes In32$\\

	\ \ \ \ $word2 \gets word1 * b $\\

    \ \ \ \ $word3 \gets (word2 \gg 22u) \otimes word2$\\ 

    $return\ Out32 \gets word3$   
\end{algorithm}
\end{small}
\end{figure}
 
 
\subsection{Chaotic behavior of our generator}
Let us recall or specify first some notations and definitions in use in this section. In what follows, $\mathbb{B}$ is the Boolean set, while $\mathbb{N}$ is the usual sets of integer numbers. For $a,b \in \mathbb{N}$, $\llbracket a, b \rrbracket$ is the set of integers: $\{a, a+1, \hdots, b\}$,
$X^\mathbb{N}$ is the set of sequences belonging in $X$ and $s^k$ is the $k$-th term of a sequence $s = \left(s^k\right)_{k\in\mathbb{N}}$, which may be a vector (thus explaining the use of an exponent). Finally, $f^n$ means the $n$-th composition of the function $f$ (\textit{i.e.}, $f^n = f \circ f\circ \hdots \circ f$). 

In the proposal, the internal function $h_f$ is iterated on the current internal state, and with a new term taken from the outer strategy. Then, the output is a permutation $\mathsf{p}$ of the internal state, which is not internally modified. 
The topological framework proposed in~\cite{secrypt16} for the CIPRNG-XOR and in~\cite{contassot2017random} can be applied, \textit{mutatis mutandis}, to this generator. It is then possible to state that iterations of the internal function are chaotic on its iteration space, denoted as
$\mathcal{X}_{32} = \mathbb{B}^{32}\times \llbracket 0, 31 \rrbracket^\mathbb{N}$. And, using a topological semi-conjugacy, that the
permutation does not alter such an unpredictable behavior. After having established that the 8-bits ICG function, denoted as $g_f$, is strongly transitive on its iteration space $\mathcal{X}_8$, we can first deduce that the discrete dynamical system $x^0 \in \mathcal{X}_8$, $x^{n+1} = g_f \left(x^n\right)$ is chaotic, and then that $h_f$ is chaotic according to Devaney.


Finally, the whole generator with the permutation $\mathsf{p}$ must be integrated inside the iterations, to see if the output has a chaotic behavior when modifying the input (internal state or strategy). To write the generator as a discrete dynamical system, we need to introduce the reverse permutation $\mathsf{p}^{-1}$.
To do so, let us define 
$$\begin{array}{lccl}
\mathsf{p}:&\mathcal{X}_{32} & \longrightarrow & \mathcal{X}_{32}\\
& (e,s) & \longmapsto & (\mathsf{p}(e),s) ,
\end{array}$$
its inverse being 
$$\begin{array}{lccl}
\mathsf{p}^{-1}:&\mathcal{X}_{32} & \longrightarrow & \mathcal{X}_{32}\\
& (e,s) & \longmapsto & (\mathsf{p}^{-1}(e),s) .
\end{array}$$
We can now introduce the following diagram:
\[
\begin{CD}
   \mathcal{X}_{32} @> h_f>> \mathcal{X}_{32}\\
   @AA\mathsf{p}^{-1}A @V\mathsf{p}VV\\
   \mathcal{X}_{32} @> G_f >> \mathcal{X}_{32}
\end{CD}
\]

$\mathsf{p}^{-1}$ and $\mathsf{p}$ are obviously continuous on $(\mathcal{X}_{32},d_{32})$, which can be directly deduced by the sequential characterization of the continuity. So the commutative diagram depicted above is a topological conjugacy, and the generator $$G_f = \mathsf{p}^{-1} \circ h_f \circ  \mathsf{p}$$ thus inherits the chaotic behavior of $h_f$ on $(\mathcal{X}_{32},d_{32})$.


\section{\uppercase{Platform and Hardware Implementation on FPGA}}
\label{sec:CIG-FPGA}

In this section, the hardware implementation of the PRNG described in this article is executed
the test platform presented in~\cite{secrypt16}. It uses a Xilinx Zynq-$7000$ 
(EPP)~\cite{rajagopalan2011xilinx} and an AXI core selector deployed as a wrapper for PRNGs. 
The main methodology of comparison between PRNGs is based on Xilinx Vivado tool $16.4$ and Zybo Board $125Mhz$.
For area comparison, we
only considered LUT and FF as follows: $$(LUT + FF) \times 8,$$ 
 Rate, for its part, is the number of bits that
are treated or transferred in each delay unit (Bps). 


Table~\ref{ICG:fpgaimp} presents the results of six different implementations of our proposal on FPGA with their TestU01 statistical test evaluations. 
During these implementations, we considered two distinct Boolean functions, namely the negation and F1 as mentioned in Section~\ref{sec:proposal}.
Three PRNGs are used as strategies (inputted generators), which are LFSR113, Taus88, and xorshift128. 
Obtained results are described hereafter.

\emph{Negation Function.} Three implementations have been realized and evaluated, see Table~\ref{ICG:fpgaimp}. 
Notice that in these 3 evaluations the value of the minimum modular multiplication operand $b$ used in the permutation (see Section~\ref{sec:proposal}) function is not the same.
To pass TestU01, it must be equal to $95$ for all strategies as LFSR113,  Tauss88, and xorshift128 (as a comparison, we found $277803737$ for PCG32). 
We have obtained that the negation function outperforms F1 in terms of throughput and area. 
Additionally, it is obviously more efficient than its best competitors recalled in this paper, as its throughput is between $1.0$ and $4$ times larger than the other chaotic PRNGs (that cannot pass TestU01), while it is $8$ times faster than the linear PRNGs... with the exception of~\cite{6868789} using the logistic map: it is true that the latter has  a throughput of $7.5$ Gbps for 32bits, but with $2$ times less area (we discarded \cite{6868789}, as this latter is fully dependent on Matlab Simulink macros, which is not relevant for ASIC implementation).
Similarly, our three implementations using the negation function exhibit less robust results compared to XOR-CIPRNG~\cite{secrypt16} for 
throughput compared to the area.
Finally, compared to the linear PRNGs that can pass TestU01 too, our three proposals with the negation function use less area and are faster. To conclude this part, and when considering the negation, our proposal  using xorshift128 as strategy is our best candidate for FPGA, and with a throughput/latency equal to $6.7$Gbps. 

\emph{F1 function.} We performed similar experiments than for the negation function. But, in these cases, the minimum modular multiplication operand $b$ is always set to $811$ for all strategies. 
We obtained a lower performance in terms of throughout when compared with the negation function, which is due to the multiplication operation in the permutation, and because in the negation we iterate a very simple logical operation (see Algorithm~\ref{eq:perm} to compare). 
However, despite its use of a bigger constant, which leads to a longer data path, the proposal with F1 does not consume any DSP block of FPGA: logic operators are sufficient. 
Additionally, results show that the three implementations with F1 function perform better than all the other chaotic PRNGs that can pass the TestU01, if we except both our proposal with the negation and the XOR-CIPRNG
Their performances are close to what has been obtained with the negation function, or to~\cite{secrypt16} with Taus88 as strategy, while F1 makes harder to reverse the process without knowing the internal transition function.

\begin{table*}[ht]
\centering
\caption{FPGA Implementation Results}
\label{ICG:fpgaimp}
\begin{adjustbox}{max width=\textwidth}
\def\arraystretch{1.0}
\begin{tabular}{|l|c|c|c|c|c|c|}
\hline
\textbf{Function}         & \multicolumn{3}{c|}{\textbf{Negation}}      & \multicolumn{3}{c|}{\textbf{F1}}            \\ \hline
PRNG                      & Taus88\_95 & LFSR113\_95 & xorshiftP128\_95 & Taus88\_811 & LFSR113\_811 & xorshiftP128\_811 \\ \hline
LUT                       & 222        & 250         & 224              & 426        & 431         & 420              \\ \hline
FF                        & 274        & 306         & 306              & 336        & 368         & 368              \\ \hline
DSP                       & 0          & 0           & 0                & 0          & 0           & 0                \\ \hline
RAM                       & 0          & 0           & 0                & 0          & 0           & 0                \\ \hline
Frequency (Mhz)           & 200        & 202       & 210.7            & 162        & 165         & 167.5              \\ \hline
Area                      & 3968       & 4448        & 4240             & 6096       & 6392        & 6304             \\ \hline
Design Latency            & 3          & 3           & 3                & 3          & 3           & 3                \\ \hline
Output Latency            & 1          & 1           & 1                & 1          & 1           & 1                \\ \hline
Throughput/Latency (Gbps) & 6.4        & 6.5         & 6.7              & 5.2        & 5.3         & 5.4              \\ \hline
\end{tabular}
\end{adjustbox}
\end{table*}

To conclude this experiment section, all what we proposed can pass all statistical tests of TestU01, from SmallCruh to BigCrush. 
Let us recall that the permutation function~\cite{citation-0} does not pass Crush and BigCrush when the space is lower than 36 bits, while in our case it does with only 32 bits and a lower modular multiplicative constant. 
These results can be improved with 64 or 128 bit outputted for a better throughput.
Finally, compared to the other CPRNG evoked in this article, we presented the only ones who can pass the stringent TestU01 battery.


\section{\uppercase{Conclusion}}
\label{sec:conclusion}
In this research work, we have introduced a new chaotic PRNG implemented in FPGA, which is based on the combination of parallel executions of generalized chaotic iterations and of an efficient permutation scheme. 
Two Boolean functions have been iterated: the vectorial negation and
one issued from removing a Hamilton cycle in the $\mathsf{N}$-cube.
Three interesting strategy builders have been evaluated for each of them. These six variations lead to an hardware generator with one of the best throughput of the literature, and that can pass the most stringent statistical batteries of tests. If we consider the two conditions of throughput and statistics, we thus have obtained one of the best existing hardware generator. 


\bigskip
\textit{This work is partially funded by the Labex ACTION program (contract ANR-11-LABX-01-01).}

\bibliographystyle{apalike}
{\small
\bibliography{References}}
\end{document}